\documentclass[prl,twocolumn,showpacs,showkeys,preprintnumbers,floatfix,superscriptaddress,footinbib]{revtex4-1}
\usepackage[colorlinks=true,linkcolor=blue,citecolor=blue,urlcolor=blue]{hyperref}
\usepackage{graphicx}
\usepackage{float}
\usepackage{xspace}
\usepackage{bm,bbm}
\usepackage{multirow,booktabs,dcolumn}
\usepackage{slashed}
\usepackage{physics}
\usepackage{nicefrac}
\usepackage[dvipsnames,usenames]{xcolor}
\usepackage[normalem]{ulem}

\usepackage{subcaption}
\newcommand{\tauv}{\bm{\tau}}

\newcommand{\mpi}{m_{\pi}}

\newcommand{\LANL}{Theoretical Division, Los Alamos National Laboratory, Los Alamos, New Mexico, 87545, USA}
\begin{document}

\preprint{LA-UR-22-25026}

\title{Auxiliary field diffusion Monte Carlo calculations of magnetic moments \\ of light nuclei with chiral EFT interactions}

\author{J.~D.~Martin}
\affiliation{\LANL}

\author{S.~J.~Novario}
\affiliation{\LANL}

\author{D.~Lonardoni}
\thanks{Present affiliation: XCP-2: Eulerian Codes Group, Los Alamos National Laboratory, Los Alamos, New Mexico 87545, USA}
\affiliation{\LANL}

\author{J.~Carlson}
\affiliation{\LANL}
  
\author{S.~Gandolfi}
\affiliation{\LANL}
  
\author{I.~Tews}
\affiliation{\LANL}

\begin{abstract}
  We calculate the magnetic moments of light nuclei ($A < 20$) using the auxiliary
  field diffusion Monte Carlo method and local two- and three-nucleon forces with electromagnetic currents from chiral effective field theory. 
  For all nuclei under consideration, we also calculate the ground-state energies and charge radii.
  We generally find a good agreement with experimental values for all of these observables.
  For the electromagnetic currents, we explore the impact of employing two different power counting schemes, and study theoretical uncertainties stemming from the truncation of the chiral expansion order-by-order for select nuclei within these two approaches.
  We find that it is crucial to employ consistent power counting schemes for interactions and currents to achieve a systematic order-by-order convergence.
\end{abstract}

\maketitle 

{\it Introduction--- }
Electromagnetic (EM) phenomena are of great importance in nuclear physics both as external probes into the structure of atomic nuclei and to understand certain internal observables.
From high-energy electron scattering that explores nuclear distributions~\cite{deforest:1966,adhikari:2021,duer:2018} to EM transition strengths that reveal details of nuclear structure~\cite{pritychenko:2017,henderson:2019,bohr:1998}, it is crucial to have a robust theoretical description of both strong and EM forces in nuclear-physics systems.
One instance of their union manifests in the magnetic moment of an atomic nucleus.
Magnetic moments are fundamental properties of nuclei which interact with atomic electrons and give rise to the hyper-fine structure in electronic spectra which can be used as a powerful tool to test quantum electrodynamics and nuclear structure. 
Additionally, nuclear magnetic moments provide a compelling test of both nuclear many-body methods and the construction of nuclear interactions and EM currents in a low-energy framework.

The nuclear magnetic moment is the vector that defines the strength and direction of the magnetic field created by an atomic nucleus. 
In a simple independent-particle model, the magnetic moment can be computed from the sum of individual nucleon magnetic moments and contributions from nonzero orbital angular momenta of protons. 
For a more realistic description, this model is greatly complicated in two ways: correlations in the nuclear wave function and inter-nucleon EM currents. 
The former can be handled by quantum many-body methods such as the quantum Monte-Carlo (QMC)~\cite{carlson:2015} method, while the latter can be handled in a consistent way by including higher-order contributions from an EM current derived from chiral effective field theory (EFT)\cite{pastore:2008,kolling:2009,pastore:2009,kolling:2011,piarulli:2013,krebs:2020}.  

\begin{figure}[!b]
    \begin{center}
        \includegraphics[width=0.47\textwidth]{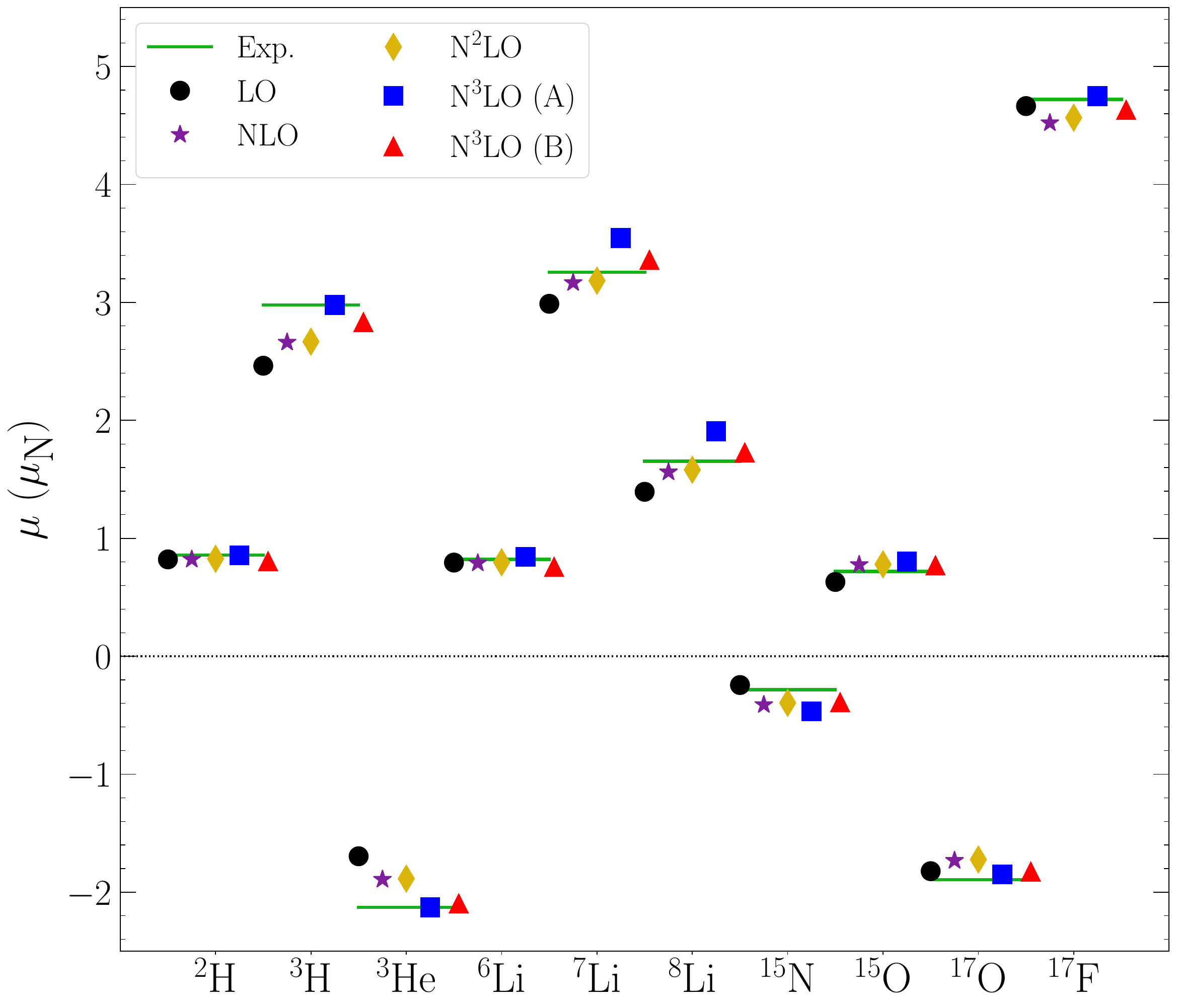} 
    \end{center}
    \caption{AFDMC results for the magnetic moments for all nuclei studied in this work.
    We show the experimental values (green lines) and results with the EM current at LO (black points), NLO (purple stars), N$^2$LO (yellow diamonds), and N$^3$LO (blue squares for scheme A and red triangles for scheme B; see text for details), while keeping the interaction fixed to be the N$^2$LO$_{\rm E\mathbbm1}$ interaction with $R_{0} = 1.0$~fm~\cite{lynn:2016}.
    The uncertainties for each experimental value and for the Monte Carlo statistics are indiscernible on this scale.}
    \label{fig:muVals_paper_pisa}
\end{figure}

Chiral EFT provides a framework for modeling the interactions both among constituent nucleons and with external probes in terms of an expansion in powers of the relevant momentum scale ($Q$) over the chiral breakdown scale ($\Lambda$). 
Retaining a finite number of terms in the chiral expansion allows one to study nuclear systems using a systematically improvable yet tractable model of the interactions.
Additionally, theoretical uncertainties can be systematically estimated by analyzing the order-by-order convergence of the expansion~\cite{Epelbaum:2014efa, Drischler:2020yad}.  
In this expansion, the lowest order term in $Q/\Lambda$ is referred to as the leading order (LO) term, the second lowest order term as next-to-leading order (NLO), the third lowest order term as next-to-next-to-leading order (N$^{2}$LO), and so on.

In this letter, we employ local interactions and EM currents from chiral EFT and use them to compute the magnetic dipole moments of light nuclei with the auxiliary field diffusion Monte Carlo (AFDMC) method~\cite{schmidt:1999,carlson:2015,lonardoni:2018b}.
In Fig.~\ref{fig:muVals_paper_pisa}, we compare the calculated magnetic moments to experimental data and generally find good agreement. 
To bolster these results, we also compute the corresponding ground-state energies and charge radii. 
We then explore different power counting (PC) schemes for the chiral expansion of the EM currents and address the consistency between these and nuclear interactions. 
Low-energy constants introduced by the EM currents at N$^3$LO are constrained by two different fits to data.
Additionally, we compute results order-by-order and use those to estimate theoretical uncertainties for both PC schemes.

{\it Methods--- }
We treat nuclei as a collection of $A$ point-like interacting nucleons of average mass $m$ described via the non-relativistic intrinsic Hamiltonian
\begin{align}
  \label{eq:hamiltonian}
  H = \sum_{i} -\frac{\nabla_i^2}{2m}+\sum_{i<j}V_{ij}+\sum_{i<j<k}V_{ijk}\,.
\end{align}
Here, the first term describes the kinetic contribution to the Hamiltonian, and $V_{ij}$ and $V_{ijk}$ are the nucleon-nucleon (NN) and three-nucleon (3N) potentials, respectively, the former of which also includes the Coulomb force. 
The NN interactions we use in this work were derived in Refs.~\cite{gezerlis:2013,gezerlis:2014} and the 3N interactions in Refs.~\cite{tews:2016,lynn:2016} and are based on a local formulation of chiral EFT.
Chiral EFT is a systematic theory for nuclear forces describing interactions in terms of a systematic momentum expansion~\cite{Epelbaum:2008ga,Machleidt:2011zz}.
It enables an improvement of interactions order-by-order and enables theoretical uncertainty estimates~\cite{Epelbaum:2014efa,Drischler:2020yad}.
The interactions employed here are derived within Weinberg PC~\cite{Weinberg:1990rz,Weinberg:1991um} but other PCs have been introduced in the past, see, e.g., Refs.~\cite{Kaplan:1996xu,Kaplan:1998tg,Kaplan:1998we,Nogga:2005hy,Long:2007vp,Long:2011xw,PavonValderrama:2014zeq,Yang:2020pgi}.
Local interactions from chiral EFT~\cite{Piarulli:2019cqu} have been successfully applied to QMC calculations of various nuclear systems~\cite{Lynn:2019rdt}.
Here, we perform AFDMC calculations of the magnetic moment using the local chiral NN and 3N interactions with cutoff $R_0=1.0$~fm up to N$^2$LO using the $E\mathbbm1$ parametrization for the 3N forces (N$^2$LO$_{\rm E\mathbbm1}$)~\cite{lynn:2016}. 
These interactions have been used to study the ground-state properties of various atomic nuclei up to $^{16}$O~\cite{lynn:2016,lynn:2017,lynn:2020,lonardoni:2018a,lonardoni:2018b,
lonardoni:2018c,cruz-torres:2019,cruz-torres:2021},
few neutron systems~\cite{klos:2016,gandolfi:2017}, and
neutron-star
matter~\cite{lynn:2016,buraczynski:2016,
buraczynski:2017,riz:2018,tews:2018a,lonardoni:2020}.


\begin{figure*}
    \centering
    \begin{subfigure}[b]{0.4\textwidth}
        \centering
        \includegraphics[width={\textwidth}]{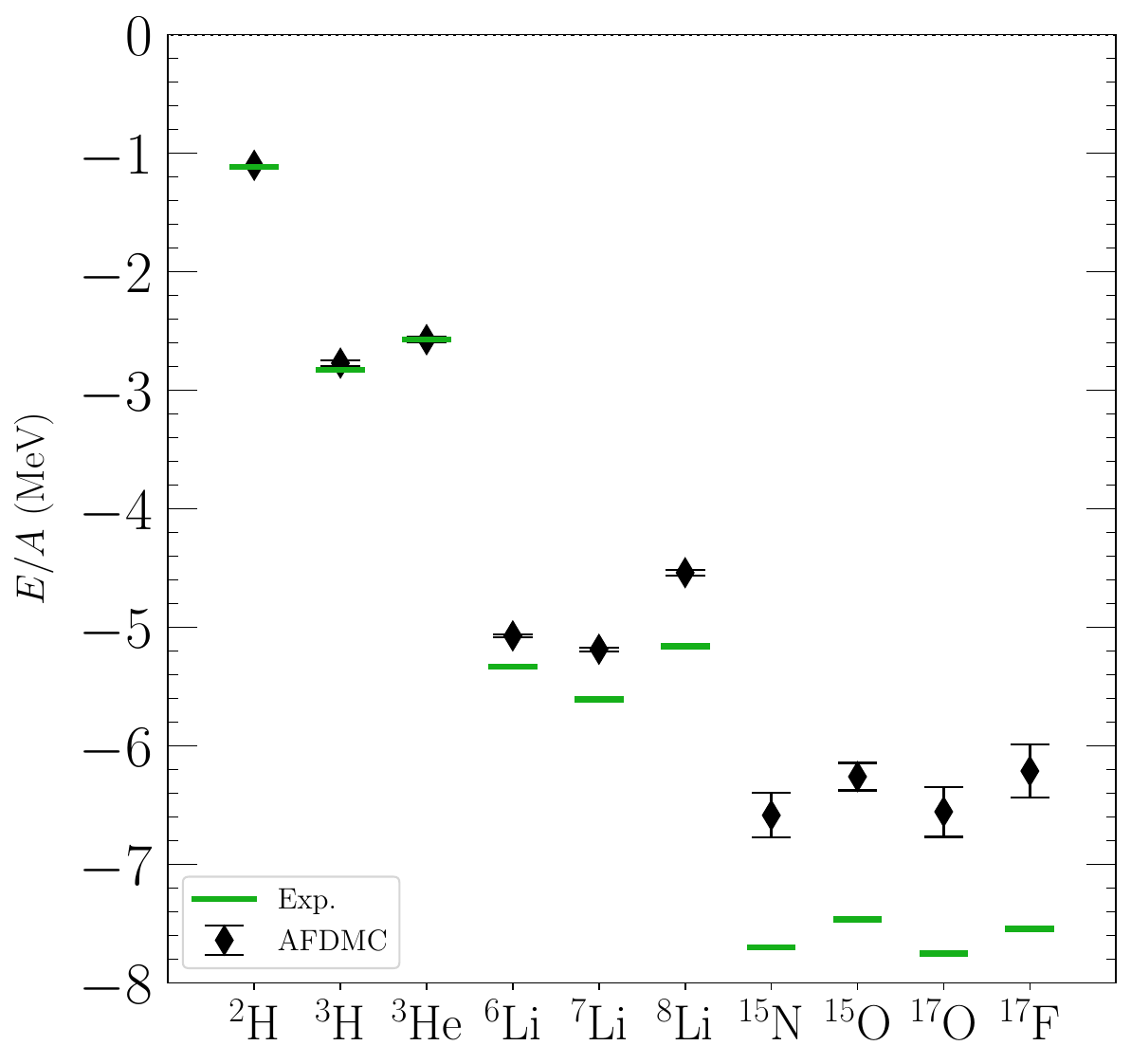}
        \caption{Binding energies per nucleon}
    \end{subfigure}
    \hspace*{0.05\textwidth}
    \begin{subfigure}[b]{0.4\textwidth}
        \centering
        \includegraphics[width={\textwidth}]{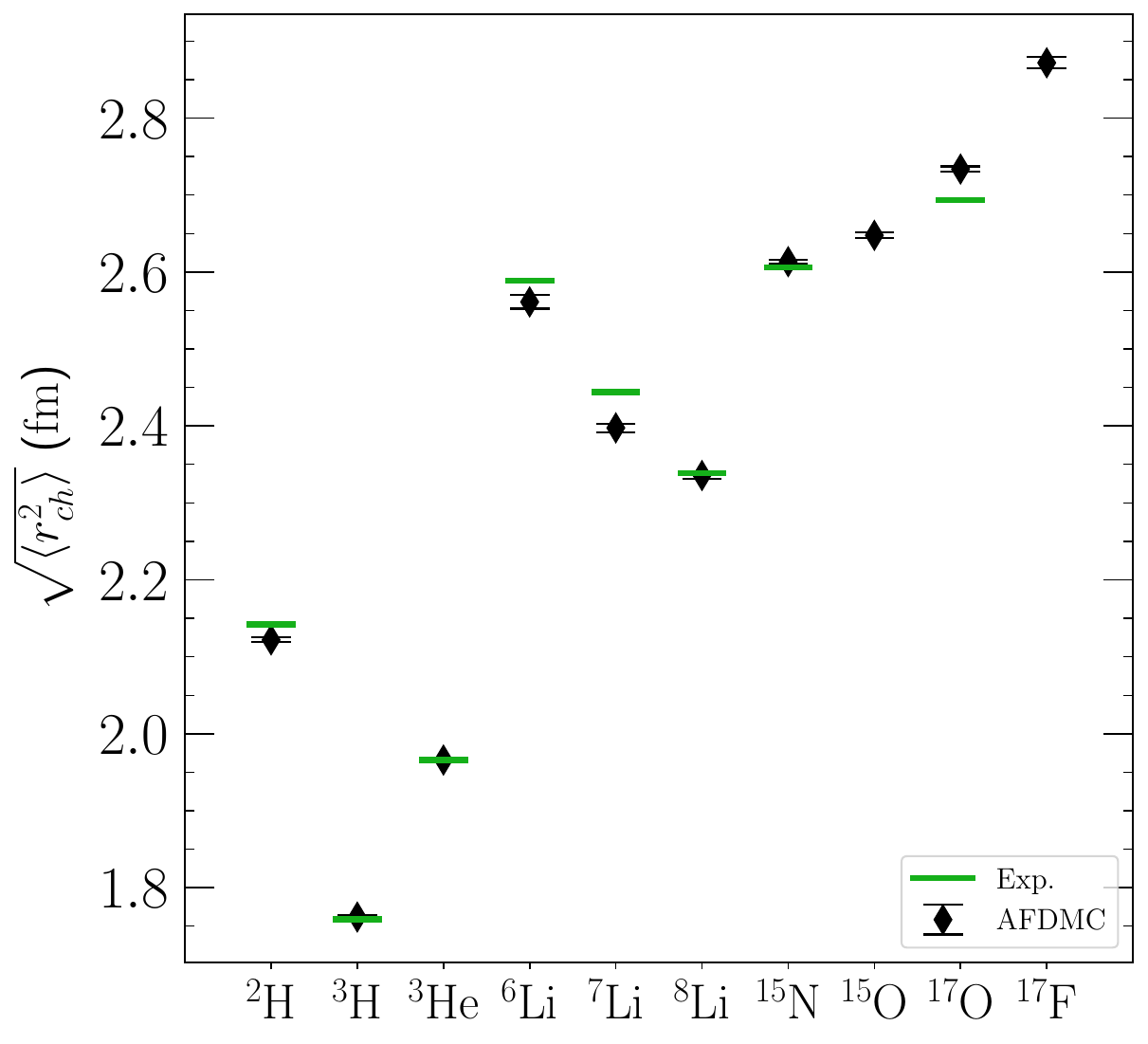} 
        \caption{Nuclear charge radii}
    \end{subfigure}
    \caption{
    Binding energies per nucleon (a) and nuclear charge radii (b) estimated using AFDMC (black diamonds) and the experimentally determined values (green lines denoted ``Exp.") for all nuclei studied in this work. Binding energies are calculated from a transient unconstrained path evolution for the N$^2$LO$_{\rm E\mathbbm1}$ interaction with $R_{0} = 1.0$ fm, while the charge radii are estimated using AFDMC using a constrained-path extrapolation. 
    (See \cite{lonardoni:2018b} for discussion on (un)constrained estimates.)
    There is no available experimental data for the charge radii of $^{15}$O and $^{17}$F. The indicated errors represent the standard error arising from the statistical uncertainty of the Monte Carlo estimate, and do not include the theoretical uncertainty arising from the truncation of the chiral expansion.
    \label{fig:nuclear_structure}}
\end{figure*}

With the Hamiltonian in hand, as the first step, we optimize the
variational trial wave function of the form
\begin{equation} \label{eq:psiTrial}
  |\Psi_{\rm T}\rangle=[F_C+F_2+F_3]|\Phi\rangle_{J^{\pi},T}\,,
\end{equation}
where $F_C$ accounts for all the spin- and isospin-independent
correlations, and $F_2$ and $F_3$ are NN and 3N correlations linear in spin- and isospin-pairs as described in Ref.~\cite{carlson:2015}. 
The term $|\Phi\rangle_{J,T}$ is taken to be
a shell-model-like state with total angular momentum $J$, parity
$\pi$, and total isospin $T$ describing the target nucleus. 
Its wave function consists of a sum of
Slater determinants ($\mathcal{D}$) constructed using single-particle orbitals:
\begin{equation} \label{eq:psiOrbitals}
  \langle RS |\Phi\rangle_{J^{\pi},T} = \sum_n c_n\left(\sum \mathcal{C}_{JT} \mathcal{D}\big\{\phi_\alpha(\vb{r}_i,s_i)\big\}\right)_{J^{\pi},T}\,,
\end{equation}
where $\vb{r}_i$ are the spatial coordinates of the nucleons and $s_i$ represent their spins.  
Each single-particle orbital $\phi_\alpha$ consists of a product of a radial function $\varphi(r)$ and an appropriate spherical harmonic coupled to the spin and isospin states.  
The determinants are coupled with Clebsch-Gordan coefficients ($\mathcal{C}_{JT}$) to total $J$ and $T$, and the $c_n$ are variational parameters multiplying different components having the same quantum numbers.
The radial functions $\varphi(r)$ are obtained by solving
for the eigenfunctions of a Wood-Saxon well, and all parameters are chosen by minimizing the variational energy as described in
Ref.~\cite{sorella:2001}. 
Then, using the AFDMC algorithm, the ground-state wave function is projected out by evolving the variational wave function with the time-propagation operator in imaginary time,
\begin{equation} \label{eq:imTProp}
  |\Psi\left(\tau\right)\rangle=\lim_{\tau\to\infty} e^{-\left(H-E_{0}\right)\tau}|\Psi_{\rm T}\rangle\ \,.
\end{equation}
More details on the AFDMC method for nuclear systems are given in Refs.~\cite{schmidt:1999,carlson:2015,Lynn:2019rdt}.

Here, we will use the AFDMC method to calculate the magnetic moments of the nuclei $^{2}$H, $^{3}$H, $^3$He, $^{6}$Li, $^{7}$Li, $^{8}$Li, $^{15}$N, $^{15}$O, $^{17}$O, and $^{17}$F. 
For these nuclei, we show the binding energies per nucleon and the charge radii without theoretical uncertainties in Fig.~\ref{fig:nuclear_structure}.
Overall, our calculations follow the trends of the experimental data. 
For the binding energies, we observe a slight underbinding consistent with the results of Ref.~\cite{lonardoni:2018a} for which all results agree with experiment within theoretical uncertainties.
For the charge radii, on the other hand, we find a very good description of experimental values.
The magnetic moment ($\mu$) is calculated from the mixed expectation value of VMC and DMC wave functions (see Ref.~\cite{lonardoni:2018b} for more details) for the magnetic form factor $F_{M}$,
\begin{align}
  F_{M}(q;\tau) \approx -i\frac{2m}{q} \Bigg[ &2\frac{\langle\Psi_{\rm T}|j_{y}(q\hat{x})|\Psi(\tau)\rangle}{\langle\Psi_{\rm T}|\Psi(\tau)\rangle} \nonumber \\
  &\; - \frac{\langle\Psi_{\rm T}|j_{y}(q\hat{x})|\Psi_{\rm T}\rangle}{\langle\Psi_{\rm T}|\Psi_{\rm T}\rangle} \Bigg] \,,
  \label{eq:mix}
\end{align}
in the limit of zero external momentum, $\mu = F_{M}(0;\tau)$.

{\it Electromagnetic current contributions --- }
We study magnetic moments by employing two different PCs for EM currents that are consistent with Weinberg PC for nuclear interactions: the Pisa~\cite{piarulli:2013,schiavilla:2019} and Bochum~\cite{krebs:2020} PCs.
In the Pisa PC, ratios of momenta to the nucleon mass are counted similar to ratios of momenta to the breakdown scale ($\Lambda_b$) so that $\frac{Q}{m} \sim \frac{Q}{\Lambda_b}$.  
The coordinate-space leading-order (LO) and next-to-leading order (NLO) contributions to the EM currents for the Pisa PC are given in Eqs.~(2.1) and (2.3) of Ref.~\cite{schiavilla:2019} and in momentum space in Ref.~\cite{piarulli:2013}.  
In contrast, the Bochum PC counts 
$\frac{Q}{m} \sim \frac{Q^2}{\Lambda_b^2}$. 
As a consequence, the lowest-order contributions to the electromagnetic vector current appear at NLO in the Bochum PC.
The total contribution to the EM current operators at NLO is identical between the Bochum and the Pisa PCs and is equal to the sum of the LO single-nucleon operators and the one-pion-exchange (OPE) NLO contributions of Ref.~\cite{schiavilla:2019}.

At N$^2$LO, there are several differences between both PCs. 
First, relativistic corrections proportional to $1/m^2$ appear at N$^2$LO in the Pisa PC while the same contribution appears at N$^{4}$LO in the Bochum PC. 
Second, the Pisa PC explicitly includes intermediate excitations of the $\Delta$ isobar.
The corresponding N$^2$LO operator is proportional to an equivalent operator contribution at N$^3$LO in a $\Delta$-less PC.  
The term at N$^{2}$LO is assumed to be significantly larger than that at N$^{3}$LO which is subsequently omitted in Ref.~\cite{schiavilla:2019}.
Because we employ a $\Delta$-less Hamiltonian, we omit this structure at N$^{2}$LO, and restore the saturated operator structure at N$^{3}$LO. 
As a result, Eq.~(2.9) of Ref.~\cite{schiavilla:2019} is modified: first, the term proportional to $\tau_{z,i}$ appears at N$^3$LO (weighted by the $\pi$N LEC $d_{8}$), and the operator proportional to $(\tauv_{i} \times \tauv_{j})_{z}$ appears at N$^{3}$LO but with a different multiplicative factor.
We therefore retain the term proportional to $(\tauv_{i} \times \tauv_{j})_{z}$ at N$^{3}$LO and make the substitution of its prefactor
\begin{equation*}
    \frac{g_{A} h_{A} \mu_{\Delta N} \mpi^{4}}{36 \pi m\,m_{\Delta N} f_{\pi}^{2} } 
        \rightarrow \frac{g_{A}^{4} \mpi^{4} }{512 \pi^{3} f_{\pi}^{4}}
\end{equation*}
We discuss our treatment of the $\Delta$-saturated N$^{3}$LO term proportional to $\tau_{i,z}$ in the next section.
Finally, we employ the commonly utilized phenomenological Sachs form factors instead of EM currents arising from the chiral expansion of these form factors which is only slowly converging \cite{phillips2016electromagnetic,schiavilla:2019,krebs:2020}.
As a consequence, there is no EM current contribution at N$^2$LO in the Bochum PC.

{\it Two nucleon, pion-nucleon, and electromagnetic LECs--- }
At N$^{3}$LO in both the Bochum and Pisa PCs, there appear three types of low-energy constants (LECs). 
First, there are NN contact LECs which already appear in the NLO nuclear interaction ($C_{2}$, $C_{4}$, $C_{5}$, and $C_{7}$).
These LECs are associated with the particular parameterization of the nuclear interaction used in Ref.~\cite{pastore:2008} which includes both local and nonlocal operators in momentum space. 
Here, instead, the Hamiltonian employed for the imaginary time propagation includes a different set of purely local operators expressed in configuration space~\cite{gezerlis:2014}, described by a different set of NN contact LECs. 
We can employ the Fierz rearrangement freedom~\cite{Huth:2017wzw} in order to relate the LECs appearing in the EM currents 
($C_i$ \footnote{See \cite{Gnech:2022vwr} for clarification on the LEC naming conventions utilized in Refs.~\cite{pastore:2008,schiavilla:2019,piarulli:2013}}) 
to our interaction LECs  ($C'_i$):
\begin{subequations}
\begin{align}
    C_{2} &= -4\,(C'_{2} + 3 \,C'_{4} + C'_{7})\,, \\
    C_{4} &= -4\,(C'_{2} - C'_{4} - C'_{7})\,, \\
    C_{5} &= -C'_{5}\,, \\
    C_{7} &= -8\,C'_{7}\,.
\end{align}
\end{subequations}
The $C'_i$ were determined from fits to NN scattering data~\cite{gezerlis:2014}.
Details on the mapping between the LECs identified via the Fierz rearrangement are provided in the Supplemental Material~\cite{supp}.

Next, there are contributions to the EM currents from the N$^3$LO pion-nucleon and N$^4$LO pion-pion Lagrangians which are proportional to the $d_{i}$ and $l_{i}$ LECs, respectively.
All except three terms proportional to the $\pi$N LECs vanish in the $q \rightarrow 0$ limit, and do not contribute to the evaluation of nuclear magnetic moments. 
The three which contribute are proportional to $d_{8}$, $d_{9}$ and $d_{21}$.  
Here, we chose to exclude all operators proportional to $d_{i}$, as these LECs are of higher order compared to the employed  
nuclear Hamiltonians.

Finally, there arise two unknown purely EM LECs that must be fit to experimental data.
These are called $L_{1}$ and $L_{2}$ in Ref.~\cite{krebs:2020} or $d_{1}^{S}$ and $d_{1}^{V}$ in Ref.~\cite{schiavilla:2019}, respectively.
We fit these two unknown EM LECs using two prescriptions, hereafter denoted scheme A and scheme B.
In scheme A, we have fit the unknown LECs to reproduce the experimental values of the magnetic moments of $^{3}$H and $^{3}$He.
For scheme B we have performed a linear least squares fit of the two LECs to the entire data set for the magnetic moments. 
Using the naming convention of Ref.~\cite{schiavilla:2019}, we find the LEC values for scheme A to be 
$d_{1}^{S} = -0.09596 \pm 0.0018$ and $d_{1}^{V} = -0.11085 \pm 0.00099$ with a $\chi^{2}$ per degree-of-freedom (DOF) of 0.48078.  For scheme B we find 
$d_{1}^{S} = -0.03958 \pm 0.0035$ and $d_{1}^{V} = -0.06420 \pm 0.0028$ with a $\chi^{2}$ per DOF of 1.5035.  

{\it Results--- }
In Fig.~\ref{fig:muVals_paper_pisa}, we show AFDMC results for the magnetic moments 
at each order in the EM current expansion as expressed in Ref.~\cite{schiavilla:2019}, omitting the terms proportional to the $d_{i}$ LECs as detailed above.
For all calculations, we fix the nuclear interaction to be the N$^2$LO$_{\rm E\mathbbm1}$ Hamiltonian and include all contributions to the EM currents up to the specified order.
None of these results include theoretical uncertainty estimates.
As a simple measure of the convergence, we consider the absolute difference between the estimated and experimental magnetic moments averaged across all nuclei at each order in the EM currents.  
This difference is 0.18 at LO, 0.13 at NLO, 0.13 at N$^2$LO, and 0.08 (0.09) at N$^3$LO in scheme A (B).
Generally, the order-by-order convergence is reasonable for all nuclei and we reproduce experimental data well.

\begin{figure}
    \begin{center}
        \includegraphics[width=0.49\textwidth]{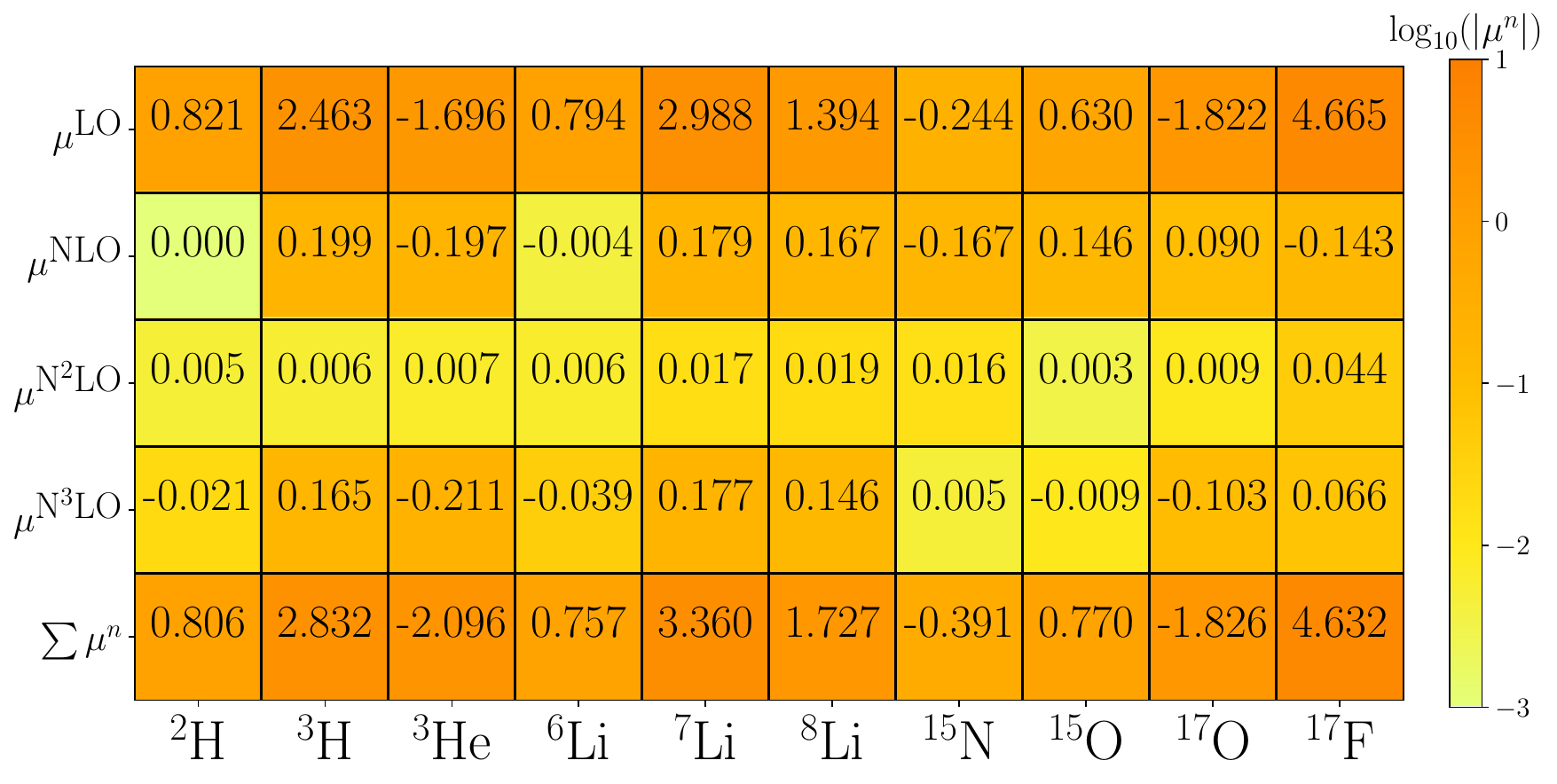} 
    \end{center}
    \caption{ Contributions to the magnetic moment at each order in the chiral expansion of the nuclear EM current for each nucleus considered. 
    We show results for the Pisa PC~\cite{schiavilla:2019} but the sum of the LO and NLO results is equivalent to the NLO contribution in the Bochum PC. 
    N$^2$LO only includes the $1/m^2$ relativistic correction while the N$^{3}$LO results are obtained by using the LECs fit in scheme B.
    The number appearing in each cell represents the MC estimate of the contribution, while the color of the cell represents the logarithm of the magnitude of the contribution to guide the eye.
    }
    \label{fig:muVals_paper}
\end{figure}

We note that N$^3$LO contributions seem to be larger than expected based on lower-order contributions, which was also observed in $A=[2,3]$ systems in Ref.~\cite{schiavilla:2019}.
To highlight this, in Fig.~\ref{fig:muVals_paper} we show the contribution of each order in the expansion for the EM currents to the total magnetic moments in the Pisa PC.
We find a nearly universal large size of the N$^{3}$LO contribution, particularly with respect to the relativistic corrections at N$^2$LO, with most nuclei displaying 
$\mathcal{O}(\mu^{\text{NLO}}) \sim \mathcal{O} (\mu^{ \text{N$^{3}$LO}})$.  
This feature is independent of the fitting scheme.

\begin{figure*} [t]
    \begin{center}
        \includegraphics[width=0.80\textwidth]{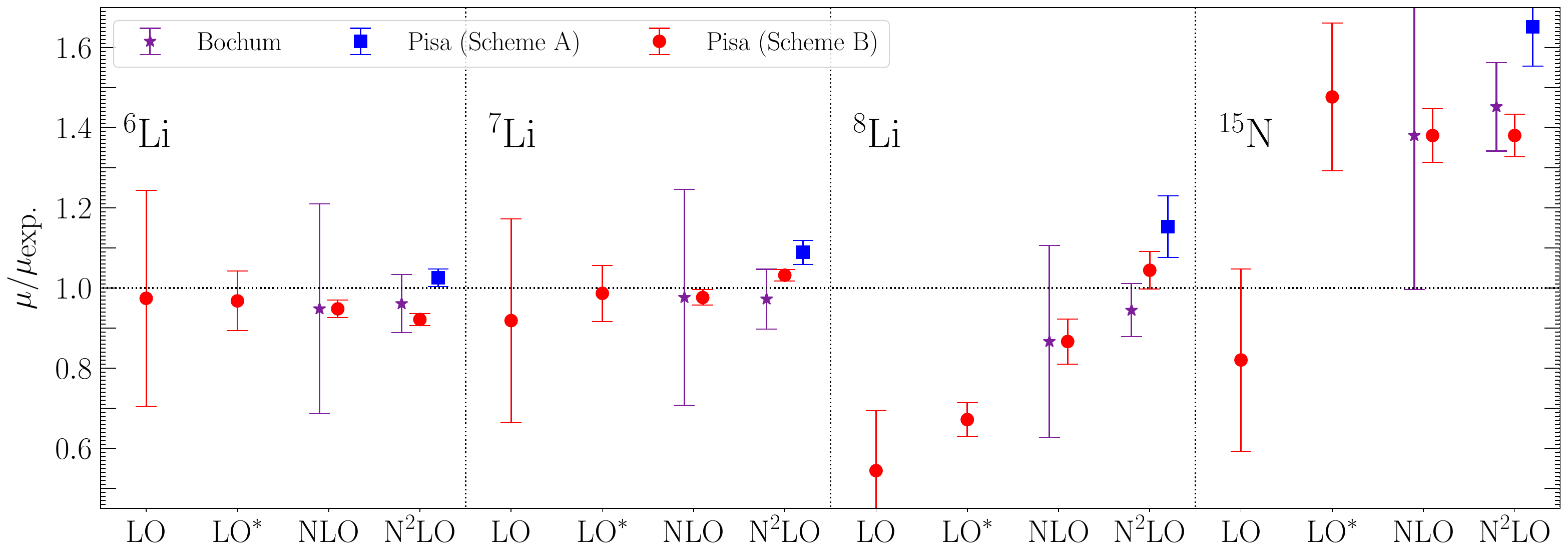} 
    \end{center}
    \caption{ 
    Order-by-order contributions to the magnetic moment relative to the experimental value for the indicated nuclei.  
    The abscissa labels indicate the order in the chiral expansion for the nuclear interaction Hamiltonian. The 
    estimates labeled $\textrm{LO}^{*}$ are calculated using the LO nuclear interactions and the NLO EM current operators.
    }
    \label{fig:uncertainties}
\end{figure*}

Next, we study the systematic uncertainty of the magnetic moments arising from the truncation of the chiral expansion in both PCs for $^{6}$Li, $^{7}$Li, $^{8}$Li, and $^{15}$N. 
For this, we perform AFDMC calculations of the magnetic moments order-by-order in the chiral expansion for \emph{both} the nuclear interaction Hamiltonian and the electromagnetic current operators, i.e., the Hamiltonian now also varies from order to order.
We use the simple prescription introduced in Ref.~\cite{Epelbaum:2014efa} and 
for the Pisa PC, the theoretical uncertainty at each order is found using the following recursive expressions:
\begin{align}
    \delta(\mu^{\text{LO}}_{\text{LO}}) &= Q | \mu^{\text{LO}}_{\text{LO}} | \,,\\
    \delta(\mu^{\text{NLO}}_{\text{LO}}) &=  \text{max} \Big\{ Q | 
        \mu^{\text{NLO}}_{\text{LO}} - \mu^{\text{LO}}_{\text{LO}} |,
        Q \delta(\mu^{\text{LO}}_{\text{LO}}) \Big\} \,,\\
    \delta(\mu^{\text{N$^{2}$LO}}_{\text{NLO}}) &= \text{max} \Big\{ Q | 
        \mu^{\text{N$^{2}$LO}}_{\text{NLO}} - \mu^{\text{NLO}}_{\text{LO}} |,
        Q \delta(\mu^{\text{NLO}}_{\text{LO}}) \Big\} \,,\\
    \delta(\mu^{\text{N$^{3}$LO}}_{\text{N$^{2}$LO}}) &= \text{max} \Big\{
    Q  | \mu^{\text{N$^{3}$LO}}_{\text{N$^{2}$LO}} - \mu^{\text{N$^2$LO}}_{\text{NLO}} |, 
        Q \delta(\mu^{\text{N$^2$LO}}_{\text{NLO}}) \Big\}\,,
\end{align}
where we take $Q = m_{\pi} / \Lambda$, and $\Lambda = 500$~MeV corresponding to our cutoff choice of $R_0=1.0$~fm~\cite{gezerlis:2014}.
In the above expressions superscripts denote the 
order in the expansion of the currents, and subscripts denote the
order in the nuclear interaction Hamiltonian.
In estimating the uncertainty, we systematically increase the order of the expansion by powers of $(Q / \Lambda)$, but because of the differences in counting powers of $1/m$ in the Pisa PC, some care is needed.  
At LO, the magnetic moment is calculated using both the Hamiltonian and the EM currents at LO in the Pisa PC.  
Because the nuclear interactions are suppressed by $(Q / \Lambda)^{2}$ at NLO relative to LO but only by $(Q / \Lambda)^{1}$ in the Pisa PC, next we  calculate the result matching LO interactions with the NLO currents. 
We include this step in our uncertainty estimate as $\delta \mu_{\textrm{LO}}^{\textrm{NLO}}$, and refer to this combination as LO$^{*}$ in Fig.~\ref{fig:uncertainties}.  
At NLO, the magnetic moments are calculated with the NLO Hamiltonian and the N$^{2}$LO currents, while at N$^{2}$LO they are calculated with the N$^{2}$LO Hamiltonian and the N$^{3}$LO currents.

Similarly, we estimate the truncation uncertainty in the Bochum PC for the EM currents, which is more consistent with the employed interaction Hamiltonian
Because the LO and N$^2$LO contributions to the magnetic moments vanish in this PC, effectively only the NLO contribution is employed.
Then, order-by-order changes in the magnetic moments arise from the nuclear interactions and the theoretical uncertainty estimate is given by 
\begin{align}
    \delta(\mu^{\text{NLO}}_{\text{NLO}}) &= Q | \mu^{\text{NLO}}_{\text{NLO}} | \,,\\
    \delta(\mu^{\text{NLO}}_{\text{N$^{2}$LO}}) &= \text{max} \Big\{ Q | \mu^{\text{NLO}}_{\text{N$^{2}$LO}} - \mu^{\text{NLO}}_{\text{NLO}} |, 
        Q \delta(\mu^{\text{NLO}}_{\text{NLO}}) \Big\}\,.
\end{align}

The ratio of the QMC estimate of the magnetic moment to the experimentally determined value for the selected nuclei is shown in Fig.~\ref{fig:uncertainties} for both the Pisa and Bochum PCs. 
For all nuclei, we observe a systematic behavior within the Bochum PC. 
However, for $^{15}$N, the calculations seem to predict a magnetic moment 40\% above the experimental value. 
For the Pisa PC, we observe a (scheme dependent) systematic convergence for $^{6}$Li, $^{7}$Li and $^{8}$Li. In contrast, for $^{15}$N the order-by-order results are less systematic.  
We also observe that for both fitting schemes, the N$^2$LO estimate lies outside the estimated error band for NLO for $^{6}$Li, $^{7}$Li and $^{8}$Li.

This highlights that consistent PCs for interactions and currents need to be employed. 
It might also point to a systematic underestimate of the characteristic momentum scale $Q$ for these nuclei, and a more detailed investigation into the one- and two-nucleon momentum distributions in these nuclei is warranted.
The choice of fitting scheme for the N$^3$LO EM LECs also produces some variance in the estimated values of the magnetic moments but this variance is of the same order as the theoretical uncertainty at NLO. 

{\it Discussion and conclusions--- }
We performed AFDMC calculations of the magnetic moments of several light nuclei with nuclear interactions and EM currents derived from chiral EFT and found agreement with experiment. 
These results are supported by comparing supplementary AFDMC calculations of the ground-state binding energies and charge-radii to experiment as well. Additionally, we explored the two prevalent PCs for the chiral EM currents and the order-by-order convergence of the magnetic moments.

While our accurate magnetic moment results are valuable in verifying the different components of these \textit{ab initio} calculations, our uncertainty analysis indicates that the consistency between employed nuclear interactions and EM currents is crucial: 
Our results for the consistent PC shows a more natural order-by-order convergence pattern.
This is also supported by the prominence of the N$^{3}$LO EM contributions to the calculated magnetic moments, and more work is necessary to study the convergence up to N$^{3}$LO in a consistent PC.
Furthermore, ensuring that the corresponding continuity equation is fulfilled is crucial~\cite{krebs:2020}. 
Future work to derive consistent currents for the interactions used in AFDMC is important.
Given the significance of uncertainty quantification for this task, it is key to properly estimate momentum scales that are being used.
Employing uncertainty quantification tools based on Gaussian Processes~\cite{Drischler:2020yad} might be beneficial for this task.

\acknowledgments{
We thank Alex Gnech, Rocco Schiavilla, Saori Pastore, Hermann Krebs, and Ronen Weiss for useful discussions.
The work of J.D.M., S.N., J.C., S.G. and I.T. was supported by the U.S. Department of Energy, Office of Science, Office of Nuclear Physics, under contract No.~DE-AC52-06NA25396. 
The work of D.L., J.C., S.G. and I.T. was 
 supported by the U.S. Department of Energy, Office of Science, Office of Advanced Scientific Computing Research, Scientific Discovery through Advanced Computing (SciDAC) NUCLEI program. 
 The work of  J.D.M., S.N. and S.G. was also supported by the Department of Energy Early Career Award Program.
The work of I.T. was also supported by the Laboratory Directed Research and Development program of Los Alamos National Laboratory under project number 20220541ECR.
Computer time was provided by the Los Alamos National Laboratory Institutional Computing Program, which is supported by the U.S. Department of Energy National Nuclear Security Administration under Contract No. 89233218CNA000001, and by the National Energy Research Scientific Computing Center (NERSC), which is supported by the U.S. Department of Energy, Office of Science, under contract No.~DE-AC02-05CH11231.
}

\bibliography{refs.bib}

\end{document}


\preprint{LA-UR-22-25026}

\title{Supplemental Material: Auxiliary field diffusion Monte Carlo calculations of magnetic moments of light nuclei with chiral EFT interactions}

\author{J.~D.~Martin}
\affiliation{\LANL}

\author{S.~J.~Novario}
\affiliation{\LANL}

\author{D.~Lonardoni}
\thanks{Present affiliation: XCP-2: Eulerian Codes Group, Los Alamos National Laboratory, Los Alamos, New Mexico 87545, USA}
\affiliation{\LANL}

\author{J.~Carlson}
\affiliation{\LANL}
  
\author{S.~Gandolfi}
\affiliation{\LANL}
  
\author{I.~Tews}
\affiliation{\LANL}

\maketitle

\section{Fierz transformation of NN contact interactions entering EM currents}

The quantum states which describe nuclei are fermionic many-body states and are therefore antisymmetric.
In chiral EFT, there are 14 sub-leading contact operators at NLO and N$^2$LO.
Because of the fermionic antisymmetry requirement, out of the 14 operators only 7 are linearly independent~\cite{gezerlis:2013}.
Hence, the choice of operators to include in the $NN$ potential is not unique, and one parameterization of the potential may be related to a different parameterization through a Fierz transformation. 
Therefore, given a particular choice of 7 operators and associated LECs, one can antisymmetrize the $NN$ potential and choose a different set of operators weighted by consistent linear combinations of the original LECs. 

The electromagnetic (EM) current operators utilized in this work were derived in chiral effective field theory from a set of nuclear interactions which implement a two-nucleon ($NN$) isospin-independent nonlocal potential at N$^{2}$LO in the chiral expansion.  
This chiral potential (Eq.~2.4 of Ref.~\cite{pastore:2009}) is 
\begin{align} \label{eq:nloCT_Pisa}
    v(\qv,\kv) = &C_{1}^{(i)} q^{2} + C_{2} k^{2} \nonumber \\
    &+ (C_{3} q^{2} + C_{4} k^{2}) \sigv_{1} \cdot \sigv_{2} \nonumber \\
    &+ \I \frac{C_{5}}{2} (\sigv_{1} + \sigv_{2}) \cdot \kv \times \qv + C_{6} \sigv_{1} \cdot \qv \sigv_{2} \cdot \qv \nonumber \\
    &+ C_{7} \sigv_{1} \cdot \kv \sigv_{2} \cdot \kv \,.
\end{align}
In this and following expressions, $\kv = (\pvec' + \pvec)/2$ and $\qv = \pvec' - \pvec$ where $\pvec$ and $\pvec'$ denote the incoming and outgoing nucleon momenta, respectively.

In contrast, the chiral potentials employed for the AFDMC imaginary time propagation in configuration space are isospin-dependent and local. 
The momentum space representation of this potential (Eq.~22 of Ref.~\cite{gezerlis:2013}) is
\begin{align} \label{eq:nloCT_Bochum}
    V(\qv,\kv) = &C_{1}' q^{2} + C_{2}' q^{2} \tauv_{1} \cdot \tauv_{2} \nonumber \\
    &+ C_{3}' q^{2} \sigv_{1} \cdot \sigv_{2} \nonumber \\ 
    &+ C_{4}' q^{2} \sigv_{1} \cdot \sigv_{2} \tauv_{1} \cdot \tauv_{2} \nonumber \\
    &+ \I \frac{C_{5}'}{2} (\sigv_{1} + \sigv_{2}) \cdot \qv \times \kv \nonumber \\
    &+ C_{6} (\sigv_{1} \cdot \qv)(\sigv_{2} \cdot \qv) \nonumber \\
    &+ C_{7} (\sigv_{1} \cdot \qv)(\sigv_{2} \cdot \qv)  \tauv_{1} \cdot \tauv_{2}\,.
\end{align}
To ensure that the correct $NN$ contact LECs enter the employed EM currents, we determine the linear transformation between the $C_{i}$ LECs appearing in the Pisa parameterization of the EM currents and their $NN$ potential, and the $C_{i}'$ LECs appearing in our interactions.

The antisymmetrization of an $NN$ potential is obtained by
\begin{equation}
    v_{\text{as}} = \frac{1}{2}(v - \mathcal{A}[v]) \,,
\end{equation}
where 
\begin{align}
    \mathcal{A}[v] = & \frac{1}{4} \left( 1 + \sigma_{12} + \tau_{12} + \sigma_{12}\tau_{12} \right) \nonumber \\
    \times &v(\qv \rightarrow -2 \kv, \kv \rightarrow -\frac{1}{2} \qv)\,,
\end{align}
and $\sigma_{12} = \sigv_{1} \cdot \sigv_{2}$ and $\tau_{12} = \tauv_{1} \cdot \tauv_{2}$.

After performing the antisymmetrization of the Pisa potential (Eq.~\ref{eq:nloCT_Pisa}) and collecting all operators, we find
\begin{widetext}
\begin{align}
    v_{as}(\qv,\kv) = \frac{1}{2} \bigg[ &\left(C_{1} - \frac{1}{16}(C_{2} 
        + 3 C_{4} + C_{7} )\right)q^{2} 
    + \left(C_{2} - (C_{1} + 3C_{3} + C_{6}) \right)k^{2} \nonumber \\
    &- \frac{1}{16}(C_{2} + 3 C_{4} + C_{7})q^{2}\tau_{12} 
        - (C_{1} + 3C_{3} + C_{6})k^{2} \tau_{12} \nonumber \\
    &+ \left(C_{3} - \frac{1}{16} (C_{2} - C_{4} - C_{7}) \right) q^{2} \sigma_{12} 
        + \left(C_{4} - (C_{1} - C_{3} - C_{6} ) \right) k^{2} \sigma_{12} \nonumber \\
    &- \frac{1}{16} (C_{2} - C_{4} - C_{7}) q^{2} \sigma_{12} \tau_{12} 
        - (C_{1} - C_{3} - C_{6}) k^{2} \sigma_{12} \tau_{12} \nonumber \\
    &+ \I \frac{3 C_{5}}{4} (\sigv_{1} + \sigv_{2}) \cdot (\kv \times \qv)
        + \I \frac{C_{5}}{4}(\sigv_{1} + \sigv_{2}) \cdot (\kv \times \qv) \tau_{12} \nonumber 
\end{align}
\begin{align} \nonumber
    &+\left( C_{6} - \frac{1}{8}C_{7} \right) (\sigv_{1} \cdot \qv) (\sigv_{2} \cdot \qv)
        + (C_{7} - 2 C_{6}) (\sigv_{1} \cdot \kv) (\sigv_{2} \cdot \kv) \nonumber \\
    &- \frac{1}{8} C_{7} (\sigv_{1} \cdot \qv) (\sigv_{2} \cdot \qv) \tau_{12} 
        - 2 C_{6} (\sigv_{1} \cdot \kv) (\sigv_{2} \cdot \kv)\tau_{12} \bigg]\, .
\end{align}
A different parameterization of the momentum-space potential is employed in the Norfolk interactions, and 
the subsequent linear mapping between the LECs relevant for the EM currents is provided in Ref.~\cite{Gnech:2022vwr}.

We similarly antisymmetrize our $NN$ potential (Eq.~\ref{eq:nloCT_Bochum}) and find
\begin{align}
    V_{as}(\qv,\kv) = \frac{1}{2}\bigg[ &C_{1}' q^{2} - (C_{1}' + 3C_{2}' + 3C_{3}' 
        + 9C_{4}' + C_{6}' + 3C_{7}' ) k^{2}  \nonumber \\
    &+ C_{2}' q^{2} \tau_{12} - (C_{1}' - C_{2}' + 3C_{3}' 
        - 3C_{4}' + C_{6}' - C_{7}') k^{2} \tau_{12} \nonumber \\
    &+ C_{3}' q^{2} \sigma_{12} - (C_{1}' + 3C_{2}' - C_{3}' - 3C_{4}' 
        - C_{6}' - 3C_{7}' ) k^{2} \sigma_{12} \nonumber \\
    &+ C_{4}' q^{2} \sigma_{12} \tau_{12} 
        - (C_{1}' - C_{2}' - C_{3}' 
            + C_{4}' - C_{6}' + C_{7}' ) k^{2} \sigma_{12} \tau_{12} \nonumber \\
    &+ \I \frac{3 C_{5}' }{4} (\sigv_{1} + \sigv_{2}) \cdot (\qv \times \kv)
        + \I \frac{C_{5}' }{4} (\sigv_{1} + \sigv_{2}) \cdot (\qv \times \kv) \tau_{12} \nonumber \\
    &+C_{6}' (\sigv_{1} \cdot \qv)(\sigv_{2} \cdot \qv) 
        - (2 C_{6}' + 6 C_{7}' ) (\sigv_{1} \cdot \kv)(\sigv_{2} \cdot \kv) \nonumber \\
    &+ C_{7}'  (\sigv_{1} \cdot \qv)(\sigv_{2} \cdot \qv) \tau_{12}
        - (2 C_{6}' - 2 C_{7}' ) (\sigv_{1} \cdot \kv)(\sigv_{2} \cdot \kv) \tau_{12} \bigg] \,.
\end{align}
\end{widetext}

We can immediately see that $C_{5}' = - C_{5}$.  To find the relationship between the other 6 LECs, we choose to match the operator set 
\begin{equation*}
\{ q^{2}, q^{2} \tau_{12}, q^{2}\sigma_{12}, q^{2} \sigma_{12} \tau_{12}, 
(\sigv_{1} \cdot \qv)(\sigv_{2} \cdot \qv),(\sigv_{1} \cdot \qv)(\sigv_{2} \cdot \qv) \tau_{12} \} \,.
\end{equation*}
From these operators, we may relate each primed LEC to a linear combination of the unprimed LECs. 
Inverting this permits us to express the unprimed LECs in terms of the primed LECs.
\begin{widetext}
\begin{equation} \label{eq:PtoG}
    \frac{1}{16}\begin{pmatrix}
        16 & -1 & 0 & -3 & 0 & -1 \\
        0 & -1 & 0 & -3 & 0 & -1 \\
        0 & -1 & 16 & 1 & 0 & 1 \\
        0 & -1 & 0 & 1 & 0 & 1 \\
        0 & 0 & 0 & 0 & 16 & -2 \\
        0 & 0 & 0 & 0 & 0 & -2 \\
    \end{pmatrix} \begin{pmatrix}
        C_{1} \\
        C_{2} \\
        C_{3} \\
        C_{4} \\
        C_{6} \\
        C_{7} \\
    \end{pmatrix} = 
    \begin{pmatrix}
        1 &  &  &  &  &  \\
         & 1 &  &  &  &  \\
         &  & 1 &  &  &  \\
         &  &  & 1 &  &  \\
         &  &  &  & 1 &  \\
         &  &  &  & & 1
    \end{pmatrix} \begin{pmatrix}
        C_{1}' \\
        C_{2}' \\
        C_{3}' \\
        C_{4}' \\
        C_{6}' \\
        C_{7}' \\
    \end{pmatrix} 
\end{equation}
\end{widetext}
Denoting the matrix in parenthesis, without the $1/16$ prefactor, as $\mathcal{M}$, Eq.~\ref{eq:PtoG} can be more compactly written as
$\mathcal{M} \vec{C} = 16 \vec{C}'$.  
Inverting this linear system permits us to write the Pisa (unprimed) LECs in terms of linear combinations of our (primed) LECs.  
For the inversion, we find that
\begin{equation}
    \mathcal{M}^{-1} = \frac{1}{16} \begin{pmatrix}
        1 & -1 & 0 & 0 & 0 & 0 \\
        0 & -4 & 0 & -12 & 0 & -4 \\
        0 & 0 & 1 & -1 & 0 & 0 \\
        0 & -4 & 0 & 4 & 0 & 4 \\
        0 & 0 & 0 & 0 & 1 & -1 \\
         0 & 0 & 0 & 0 & 0 & -8
    \end{pmatrix} .
\end{equation}

Thus, we obtain
\begin{equation} \label{eq:PtoG2}
    \begin{pmatrix}
        C_{1} \\
        C_{2} \\
        C_{3} \\
        C_{4} \\
        C_{6} \\
        C_{7} \\
    \end{pmatrix} = 
    \begin{pmatrix}
        1 & -1 & 0 & 0 & 0 & 0 \\
        0 & -4 & 0 & -12 & 0 & -4 \\
        0 & 0 & 1 & -1 & 0 & 0 \\
        0 & -4 & 0 & 4 & 0 & 4 \\
        0 & 0 & 0 & 0 & 1 & -1 \\
         0 & 0 & 0 & 0 & 0 & -8
    \end{pmatrix} \begin{pmatrix}
        C_{1}' \\
        C_{2}' \\
        C_{3}' \\
        C_{4}' \\
        C_{6}' \\
        C_{7}' \\
    \end{pmatrix} .
\end{equation}
We have further validated that different choices for the set of operators in the LEC matching procedure leads to the same mapping between LECs for the two different parameterizations of the N$^{2}$LO chiral potentials.

\section{Power counting comparison}

The power counting (PC) schemes employed by the Pisa and Bochum groups in their respective derivation of the electromagnetic currents differ in how they count inverse 
powers of the nucleon mass. In table \ref{tab:PCs} we show the highest order included at each order in the chiral expansion for the Hamiltonian we employ, as well 
as the expansion order for the current operators derived by the Pisa and Bochum groups.

\begin{table} [!h]
    \centering
    \begin{tabular}{| c | c | c | c |}
        \hline
        Order & $H$ & $J_{\textrm{Pisa}}$ & $J_{\textrm{Bochum}}$ \\
        \hline
        \hline
        LO & $Q^{0}$ & $Q^{-2}$ &  $Q^{-3}$  \\
        \hline 
        NLO & $Q^{2}$ & $Q^{-1}$ & $Q^{-1}$ \\
        \hline
        N$^{2}$LO & $Q^{3}$ & $Q^{0}$ & $Q^{0}$ \\
        \hline
        N$^{3}$LO & $Q^{4}$ & $Q^{1}$ & $Q^{1}$ \\
        \hline
    \end{tabular} 
    \caption{Highest order term in the chiral expansion of the Hamiltonian employed in this work ($H$ column), the EM current operators derived by the Pisa group ($J_{\textrm{Pisa}}$ column),
        and the EM current operators derived by the Bochum group ($J_{\textrm{Bochum}}$). Note that although LO is defined at $\mathcal{O}(Q^{-3})$ in the Bochum PC, 
        there is no contribution to the vector current operator at this order in this PC.}
    \label{tab:PCs}
\end{table}

The total summed contribution to the EM currents is identical between the Pisa and Bochum PC's.  However, the LO contribution to the EM vector current is 0 in the Bochum PC.  This is because the term which appears at LO in the Pisa PC is proportional to $Q / m$ (the ratio of momentum to the nucleon mass) which is counted as one order higher in the Bochum PC.

\section{Individual contributions to the EM current}

\begin{figure} [!h]
    \begin{center}
        \includegraphics[scale=0.27]{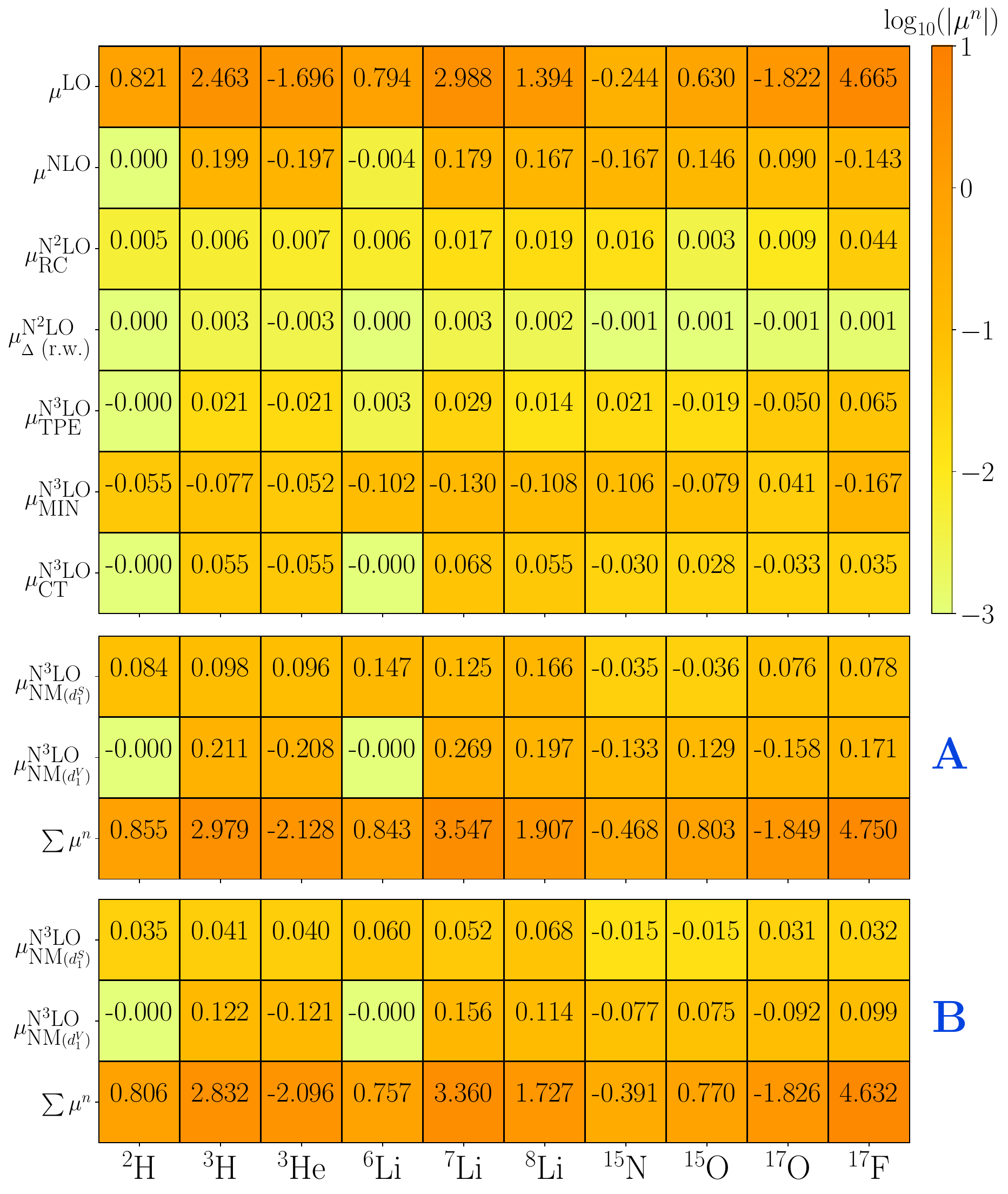}
    \end{center}
    \caption{Evaluated contribution to the magnetic moment for each term in chiral expansion of the EM current operator at N$^{2}$LO in the Pisa PC. The top 
    panel are the values common to both EM LECs fitting schemes.  The first two rows of the middle panel are the contributions to the magnetic moment from the terms proportional to the unknown 
    EM LECs fit with scheme A, and the final row is the total evaluated magnetic moment including all terms of the first panel and first two rows of the middle panel.  The bottom panel 
    is similar to the middle panel, but the top two rows represent those terms proportional to the unknown EM LECs fit with scheme B.}
    \label{fig:all_currents}
\end{figure}

\bibliography{refs.bib}